\documentclass[aps,prl,twocolumn,groupedaddress,showpacs]{revtex4-1}
\usepackage{graphicx}
\usepackage{psfrag}
\def\be{\begin{equation}}
\def\ee{\end{equation}}

\begin{document}
\date{\today}

\title{Symmetry Breaking in d-Dimensional Self-gravitating Systems}
\author{Renato Pakter$^1$}
\author{Bruno Marcos$^{1,2}$}
\author{Yan Levin$^1$}
\affiliation{$^1$Instituto de F\'{\i}sica, Universidade Federal do Rio Grande do Sul, Caixa Postal 15051, CEP 91501-970,
Porto Alegre, RS, Brazil\\ $^2$Universit\'e de Nice Sophia-Antipolis, CNRS, Laboratoire J.-A. Dieudonn\'e, UMR 7351, Parc Valrose,  06108 Nice Cedex 02, France.}

\begin{abstract}

Systems with long-range interactions, such as self-gravitating clusters and magnetically 
confined plasmas, do not relax to the usual Boltzmann-Gibbs thermodynamic
equilibrium, but become trapped in quasi-stationary states (QSS) the life
time of which diverges with the number of particles. The QSS are characterized by the lack of
ergodicity which can result in a symmetry broken QSS starting from a spherically symmetric
particle distribution. We will present a theory which allows us to 
quantitatively predict
the instability threshold for spontaneous symmetry breaking for   
a class of $d$-dimensional self-gravitating systems.

\end{abstract}


\maketitle

Lord Rayleigh was probably the first to make an observation that long-range forces 
can lead
to symmetry breaking~\cite{Re82}.  Rayleigh was studying stability of conducting spherical fluid droplets
carrying charge $Q$.  He discovered that when $Q$ exceeds a certain critical threshold $Q_c$,
droplets becomes unstable to symmetry breaking perturbations, elongating and eventually
breaking up, emitting jets of fluid that carry away a significant fraction of the  
charge~\cite{Ta64}. Rayleigh instability is now the basis for various technological applications, such as 
electrospraying
and electrospinning.   It also helps to understand the conformational structure of charged polymers, such
as polyampholytes~\cite{KaKa95}.
For self-gravitating systems a similar instability has been observed in gravitational simulations~\cite{AgMe90}.  
It has been found that an initially spherically symmetric self-gravitating system 
can become unstable, leading to formation of structures of reduced
symmetry~\cite{AgMe90}.  This radial orbit instability is believed to be important for 
the formation of elliptical galaxies~\cite{AtMa13}.

There is, however, a fundamental difference between the Rayleigh
instability of charged spherical droplets and the instability of  spherically symmetric 
self-gravitating systems.
Since the droplets are in (canonical) thermodynamic equilibrium, their shape must correspond to the 
minimum of 
the Helmholtz free energy  --- in fact, even for $Q$ somewhat below $Q_c$
a spherical shape is already metastable, with the global minimum corresponding to a strongly prolate ellipsoid~\cite{AiGa62}.  
The thermal fluctuations, however, are too small  
to overcome the barrier that separates the metastable minimum from the 
global one, so that the spherical shape persists up to the Rayleigh threshold.
On the other hand, gravitational systems are intrinsically microcanonical --- isolated from
environment~\cite{Gr01,IsCo01,ChIs02}. In the thermodynamic limit,  such long-range systems do
not evolve to thermodynamic equilibrium, but become trapped in quasi-stationary states (QSS),
the life time of which
diverges with the number of particles~\cite{Campa09}. The QSS are characterized by the broken ergodicity,
making equilibrium statistical mechanics inapplicable~\cite{BeTe12}. 
To explore spontaneous symmetry breaking of systems
with long-range forces, therefore, requires a completely different approach~\cite{TeBe12}.   
In this Letter we will present a theory which allows us to quantitatively predict the thresholds
of symmetry breaking instabilities for systems with long-range interactions.  
The results of the theory
will be compared with extensive molecular dynamics simulations.  

To present the theory, we will study a class of self-gravitating systems of 
$N$ particles of mass $m$ in an infinite $d$-dimensional space. The interaction potential between
the particles is $V({\bf r})=\frac{Gm^2}{(2-d) r^{d-2}}$, where $G$
is the gravitational constant.
We will work in thermodynamic limit, 
$N \rightarrow \infty$ and $m \rightarrow 0$, while the total mass $M\equiv Nm$ remains fixed.  The initial 
particle distribution is assumed to be a uniform spherically symmetric 
water-bag in both configuration and velocity space, 
\be
f_0({\bf r},{\bf v})=\frac{d^2}{C_d^2 r_m^d v_m^d} \Theta(r_m-r) \Theta(v_m-v),
\label{f0}
\ee
where $\Theta$ is the Heavyside step function and $C_d=2\pi^{d/2}/\, \Gamma(d/2)$ is the surface area of a 
$d$-dimensional unit sphere, and $\Gamma(x)$ is
the gamma function. Since the initial water-bag distribution is not a stationary solution of the 
collisionless Boltzmann (Vlasov) equation the systems will evolve with time. 
We are interested to discover under what conditions Eq.(\ref{f0}) becomes unstable
to small non-axisymmetric perturbations.

It is convenient to define dimensionless variables by scaling the
distance, time, velocities, the gravitational potential, and the energy with respect to: $r_0=r_m$, 
$t_0=\sqrt{r_m^d/GM}$, 
$v_0 =\sqrt{GM/r_m^{d-2}}$, $\psi_0 = GM/r_m^{d-2}$ and $E_0 =GM^2/r_m^{d-2}$, respectively.
This is equivalent to setting $r_m=G=M=1$.
The particle dynamics is governed by Newton's equations of motion 
\be
\ddot {\bf r}=-\nabla \psi({\bf r},t),
\label{eqmotion}
\ee
where the dot
stands for the time derivative and ${\bf r}=\sum_i x_i\hat{\bf e}_i$, $i=1 \cdots d$, is the particle position.  
In thermodynamic limit, the correlations between the particles can be 
ignored, so that the force acting on a particle located at ${\bf r}$ is ${\bf F}=-\nabla \psi({\bf r}, t)$, where
$\psi({\bf r}, t)$  is the mean gravitational potential which satisfies the Poisson equation
\be
\nabla ^2 \psi=C_d\, n({\bf r},t),
\label{poisson}
\ee
where $n({\bf r},t)$ is the particle number density. 

We define the "envelope" of the position and velocity particle distributions to be 
$X_i(t)=\sqrt{(d+2)\langle x_i^2\rangle}$ and 
$V_i(t)=\sqrt{(d+2)\langle v_i^2\rangle}$, respectively.
The $\langle \cdots\rangle$
corresponds to the average over all the particles. 
Note that in the reduced units, $X_i(0)=1$ and $V_i(0)=v_m$ for all $i$, but
as the dynamics evolves, it is possible for the symmetry between the different directions to become broken. 
Our goal is to determine the equations of evolution for $X_i(t)$ \cite{Sa71}.
Taking two time derivatives of $X_i^2(t)$ and one of $V_i^2(t)$ and using the 
equations of motion, Eq.~(\ref{eqmotion}), we obtain
\be
\dot X_i^2 +X_i \ddot X_i = V_i^2-(d+2) \left \langle x_i {\partial \psi\over \partial x_i}\right \rangle
\label{pos}
\ee
and
\be
V_i \dot V_i = -(d+2)\left \langle \dot x_i {\partial \psi\over \partial x_i}\right \rangle \,.
\label{vel}
\ee

To calculate the averages appearing in Eqs. (\ref{pos}) and (\ref{vel}), 
we need to know the mean-gravitational potential.  
We suppose 
that the originally spherically symmetric homogeneous distribution can become
distorted into an ellipsoidal shape with
the semi-axis $\{ X_i\}$ and 
uniform density $n({\bf r}, t)=d/C_d\prod_i X_i(t)$. Using the ellipsoidal coordinate system~\cite{Ke53} the
gravitational field inside a $d$-dimensional ellipsoid with the semi-axis $\{ X_i\}$ 
can be calculated explicitly to be,
\be
\frac{\partial \psi}{\partial x_i}=\frac{d}{2} x_i g_i(X_1,\cdots,X_d),
\label{field}
\ee 
where
\be
g_i(X_1,\cdots,X_d)=\int_0^\infty {ds\over (X_i^2+s)\prod_{j=1} ^d (X_j^2+s)^{1/2}}.
\label{gi}
\ee
Furthermore, for a $d$-dimensional ellipsoid with a uniform
mass distribution it can be shows that $\langle x_i^2 \rangle = X_i^2/(d+2)$.
Substituting these results in  
Eqs. (\ref{pos}) and (\ref{vel}), we obtain a closed set of 
coupled equations 
\be
\dot X_i^2 +X_i \ddot X_i = V_i^2-{d\over 2} X_i^2 \, g_i(X_1,\cdots,X_d)
\label{pos1}
\ee
and
\be
V_i \dot V_i = {d\over 2} X_i \dot X_i \, g_i(X_1,\cdots,X_d) \,.
\label{vel1}
\ee
We define the "emittance" in the $i$'th direction as
$\epsilon_i^2(t)\equiv (d+2)^2[\langle x_i^2\rangle \langle \dot x_i^2\rangle-\langle x_i \dot x_i\rangle ^2]$=$X_i^2 V_i^2-\dot X_i^2 X_i^2$.
Taking a time derivative of $\epsilon_i^2(t)$ and using the equations (\ref{pos1}) and (\ref{vel1}), it
is possible show that the  $\epsilon_i(t)$ are the constants of motion, 
$\epsilon_i(t)=\epsilon_i(0)\equiv \epsilon_i$. 
Using this observation the set of Eqs (\ref{pos1}) and (\ref{vel1}) reduces to
\be
\ddot X_i={{\epsilon_i^2}\over X_i^3}-{d\over 2}\, X_i\, g_i(X_1,\cdots,X_d).
\label{enve2} 
\ee
For the initial water-bag distribution, Eq. (\ref{f0}), $\epsilon_i^2(0)=v_m^2$. 

The virial theorem requires that a {\it stationary}
gravitational system in $d$ dimensions must have
$2K=(2-d)U$, where $K$ and $U$ are the total kinetic and
potential energies, respectively. 
For the initial water-bag distribution $K=\frac{ v_m^2 \, d}{2 (d+2)}$ and
the potential energy is $U=\frac{d}{(2-d)(d+2)}$, so that the virial condition reduces to
$v_m=1$. Although the initial water-bag distribution is not a stationary solution of the collisionless Boltzmann (Vlasov) equation, we expect that if the virial condition is satisfied, the system will 
not exhibit
strong envelope oscillations. This is indeed what has been observed for gravitational systems in 
$d=1,2$ and $3$~\cite{TeLe11,JoWo11,TeLe10,LePa08a}. 
On the other hand if the initial distribution does not satisfy the virial condition, the
particle distribution will undergo violent oscillations which will lead to QSS with a core-halo 
structure~\cite{TeLe11,TeLe10}.  
To measure how strongly the initial distribution
deviates from the virial condition, we define a viral number ${\cal R}_0\equiv \frac{2K}{(2-d)U}=v_m^2$.    
With this definition the emittance becomes $\epsilon_i^2(t)={\cal R}_0$.

Let us first consider a uniform spherically symmetric mass distribution 
of radius $R(t)$, i.e., $X_i(t)=R(t)$ for $i=1,\cdots,d$. In this case the integral in
Eq.~(\ref{gi}) can be evaluated analytically to yield $g_i=2R^{-d}/d$, and the 
equation of evolution for the radius of the sphere becomes
\be
\ddot R={{\cal R}_0\over R^3}-{1\over R^{d-1}},
\label{R} 
\ee
with $R(0)=1$ and $\dot R(0)=0$. We see that in agreement with the earlier discussion, 
if the initial distribution satisfies the virial condition, ${\cal R}_0=1$, 
the sphere's radius remains constant for all time, $R(t)=1$  for any $d$. 
For $d\leq 3$, this equilibrium
is stable because a small deviation from ${\cal R}_0=1$  will result in small periodic
oscillations of $R$. On the other hand, for $d\geq 4$ the equilibrium is unstable, and any
${\cal R}_0\not =1$ will lead to either collapse or an unbounded expansion of the particle distribution.
These conclusions are in agreement with the old observation of Paul Ehrenfest, who first noted 
that there are no stable orbits for Newtonian gravity in $d\geq 4$~\cite{Eh20}.

To investigate the possible symmetry breaking of an initially spherically symmetric mass distribution 
we need, therefore, to only consider $d=2$ and $3$. For $d=2$,
the integral in Eq.~(\ref{gi}) can be performed analytically yielding $g_i(X_1,X_2)=2/X_i(X_1+X_2)$.
Eq. (\ref{enve2}) then simplifies to
\be
\ddot X_i={{\epsilon_i}\over X_i^3}-{2\over X_1+X_2},\,\,\, i=1,2.
\label{enve2d} 
\ee
The symmetry breaking occurs if an initially vanishingly small fluctuation
grows as a function of time.
To study this instability, it is convenient to introduce new variables
\be
X_i(t)=\bar{X}(t)+\Delta_i(t),
\label{defs}
\ee
where $\bar{X}=(\sum_i X_i)/d$ is the average of $X_i$'s and $\Delta_i$ is the asymmetry along the $i$th 
direction.
Clearly  $\Delta_i$'s are related by $\sum_i \Delta_i=0$. Hence, for $d=2$ there is only one independent 
asymmetry variable
$\Delta=\Delta_1=-\Delta_2$.  
To locate the region of instability, we perform a linear stability analysis of Eqs. (\ref{enve2d}). 
Noting that $(\epsilon_1^2 - \epsilon_2^2) \sim O(\Delta)$,
to leading order in $\Delta$,  Eq. (\ref{enve2d}) simplifies to
\be
\ddot \Delta + {3 (\epsilon_1^2 + \epsilon_2^2)\over 2 \bar X^4(t)}\, \Delta= {(\epsilon_1^2 - \epsilon_2^2)\over 2 \bar X^3(t)},
\label{Delta}
\ee
while the dynamics of $\bar X(t)$ to this order is
\be
\ddot {\bar X}={\epsilon_1^2+\epsilon_2^2\over 2 \bar X^3}-{1\over \bar X}.
\label{barx} 
\ee
The dynamics of $\Delta$ is driven by the oscillations of $\bar X(t)$.
In particular, if the virial condition is satisfied and $\epsilon_1^2 =\epsilon_2^2={\cal R}_0=1$,
the  $(\Delta=0,\dot\Delta=0)$ is a stable fixed point of Eq. (\ref{Delta}).   
Therefore if ${\cal R}_0 \approx 1$, for small initial asymmetry, $\Delta(t)$ will not grow in time. 
However, if the initial distribution does not satisfy the virial condition, $\bar X(t)$ will 
oscillate and may drive a parametric resonance which can make $\Delta(t)$  unstable. 
This is precisely what is observed in numerical integration of Eqs. (\ref{Delta}) and (\ref{barx}). We 
find that for sufficiently small (or large)  ${\cal R}_0$, the amplitude of $\Delta(t)$ oscillations
grows without a bound.  Note that in Eq. (\ref{Delta}) the instability occurs as a consequence
of a fluctuation either in the velocity ($\Delta(0)=0$ and  $\epsilon_1 \neq \epsilon_2$), the
position ($\Delta(0) \neq 0$), or as a combination of both. For sufficiently 
small (or large)  ${\cal R}_0$, we find that any small fluctuation in the initial particle
distribution is amplified by the dynamics.  Of course, in practice the growth of $\Delta(t)$ will be saturated
by the Landau damping~\cite{TeLe11,TeLe10} and will result in a QSS with a broken rotational symmetry.
   
To precisely locate the instability threshold it is simplest to consider a small fluctuation with 
$\Delta(0) \neq 0$ and $\epsilon_1 = \epsilon_2$.
Since the $\Delta(t)$ is driven by the periodic oscillations of $\bar X(t)$, to study this instability 
we must work in the Poincar\'e section~\cite{PoHi93,SiRe06}. 

\begin{figure}[h]
\psfrag{X}{$t$}
\psfrag{Y}[c]{$\langle x_i^2 \rangle$}
\psfrag{Z}{${\cal R}$}
\includegraphics[height=4.1cm]{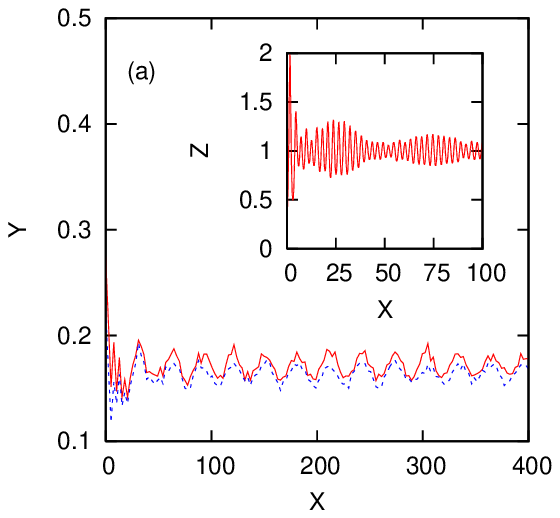}
\hspace{-0.6cm}
\psfrag{Y}{}
\includegraphics[height=4.1cm]{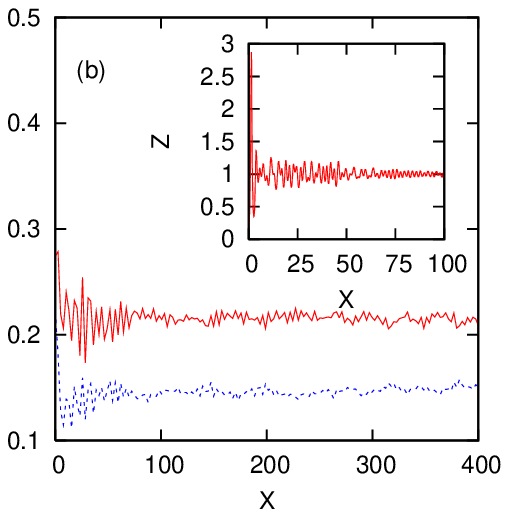}
\caption{The evolution of $x$ and $y$ moments of the mass distribution, 
$\langle x^2 \rangle$ (solid curve, red) and $\langle y^2 \rangle$ (dashed curve, blue) 
obtained using molecular dynamics 
simulations for a 2D system with $N=8000$. 
A small asymmetry in the $x$-direction is introduced in the initial particle distribution.
For initial distribution with ${\cal R}_0=0.36$ (panel (a) ) 
the system relaxes to a QSS with a spherical symmetry(see also Fig. \ref{fig2b}), 
while for ${\cal R}_0=0.16$ (panel (b) ) spherical symmetry is broken.  Similar behavior is 
found for  ${\cal R}_0$ above the upper critical threshold, see Fig. \ref{fig2b}. The inset shows the 
evolution of the virial number.  Both the symmetric and the asymmetric QSS are fully virialized, ${\cal R}=1$}\label{fig2}
\end{figure}

\begin{figure}[h]
\psfrag{X}{$x$}
\psfrag{Y}{$y$}
\includegraphics[height=4.0cm]{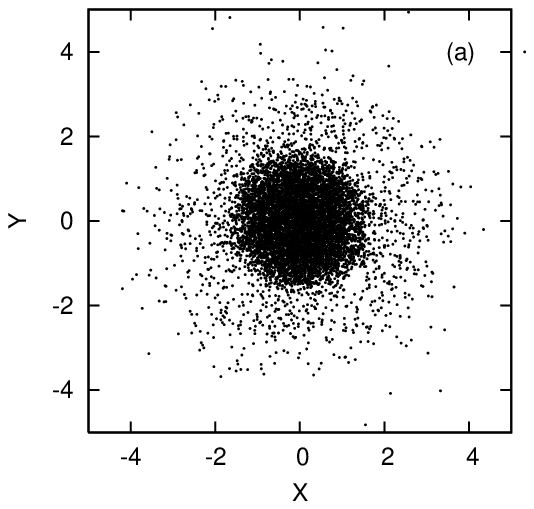}
\psfrag{Y}{}
\hspace{-0.6cm}
\includegraphics[height=4.0cm]{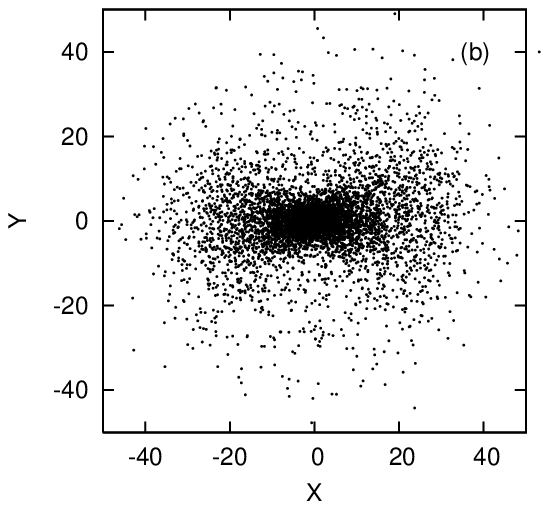}
\caption{Snapshots of the $x-y$ particle distribution in a QSS at t=200.  
In panel (a) ${\cal R}_0=2$ and the symmetry remains unbroken, while in panel (b) ${\cal R}_0=6.25$ and the QSS
has a broken rotational symmetry.  Note that the final particle distribution in both
cases has a characteristic core-halo structure.}\label{fig2b}
\end{figure}

Consider a displacement vector from the 
$(\Delta=0,\dot\Delta=0)$ fixed point, ${\bf X}_\Delta(t)=(\delta\Delta,\delta \dot{\Delta})$. 
From Eq.~(\ref{Delta}), we see that its dynamics is governed by 
$\dot {\bf X}_\Delta={\bf M}\cdot {\bf X}_\Delta$, where
\be
{\bf M}=\left( \begin{array}{cc}
0 & 1 \\
-{3{\cal R}_0\over \bar X^4} & 0\end{array} \right),
\ee
and the dynamics of $ \bar X(t)$ is given by Eq.~(\ref{barx}) with $\epsilon_1^2=\epsilon_2^2={\cal R}_0$. 
If we now define a mapping  ${\cal M}(t)$ that relates ${\bf X}_\Delta(t)$ to its initial condition by 
${\bf X}_\Delta(t)={\cal M}(t)\cdot{\bf X}_\Delta(0)$, and substitute this into 
the evolution equation for ${\bf X}_\Delta$,
we obtain 
\be
\dot {\cal M}={\bf M}\cdot {\cal M},\, {\rm with}\, {\cal M}(0)=I,
\label{m}
\ee
where $I$ is the identity matrix. In order to determine the stability of $(\Delta=0,\dot\Delta=0)$ fixed point, 
we simultaneously integrate Eqs. (\ref{R})
and (\ref{m}) over one period $\tau_R$ of the oscillation of $\bar X(t)$  (i.e., between two consecutive points in the 
Poincar\'e map), and determine the eigenvalues 
of the mapping matrix ${\cal M}(\tau_R)$. 
If the absolute value of any eigenvalue is larger than 1, then $(\Delta=0,\dot\Delta=0)$ fixed point will 
be unstable. We 
find that the asymmetric instability occurs for ${\cal R}_0<0.255893...$ and for ${\cal R}_0>2.55819...$. 
A more detailed analysis shows that it is produced by a pitchfork bifurcation and is of 
second order. In Fig. \ref{fig2} we compare the predictions of the theory with the results of extensive
molecular dynamics simulations performed using the state-of-the-art gravitational oriented massively parallel 
GADGET2 code~\cite{Gad}, which has been appropriately  modified to integrate gravity in two dimensions.  
At $t=0$ the particles are distributed in accordance with
Eq. (\ref{f0}).  To force the symmetry breaking to occur along the $x$-axis, a small perturbation in this direction 
is introduced.  We then monitor the moments $\langle x^2 \rangle$  and $\langle y^2 \rangle$ as the dynamics evolves.  
Fig. \ref{fig2} shows the evolution of the moments for two different virial numbers.  
We find that for ${\cal R}_0=0.16$ 
the symmetry is broken while for ${\cal R}_0=0.36$
the spherical 
symmetry is unaffected by the initial
perturbation.  This is in close agreement with the predictions of the present theory.  
Similar symmetry breaking transition is also found for large virial numbers, see Fig. \ref{fig2b}. Since the 
transitions are 
continuous, it is difficult to precisely locate the thresholds of instability 
using molecular dynamics simulations.  

\begin{figure}[h]
\psfrag{X}{x}
\psfrag{Y}[c]{$y$}
\includegraphics[height=4cm]{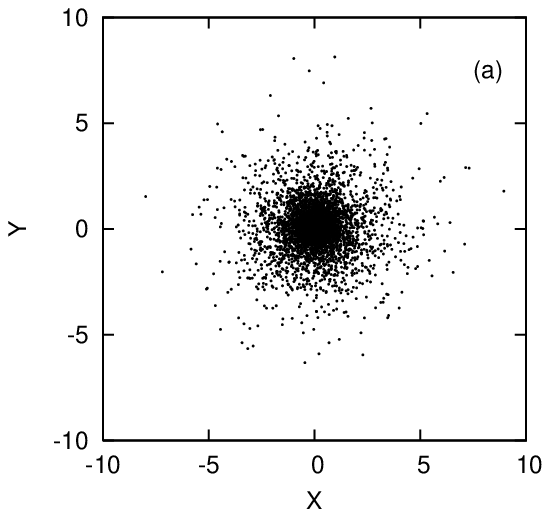}
\hspace{-0.3cm}
\psfrag{Y}{}
\includegraphics[height=4cm]{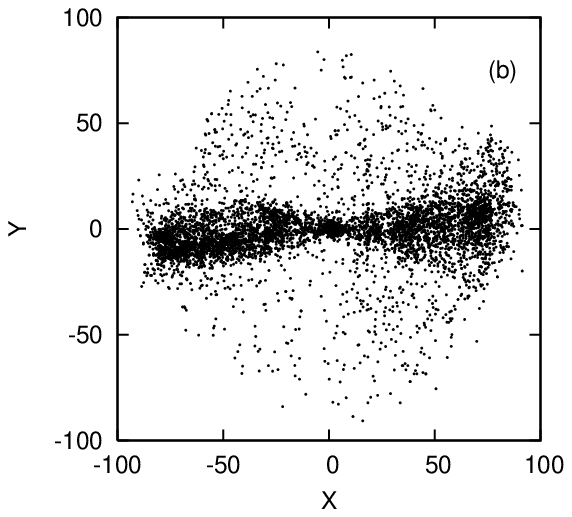}
\caption{Snapshots of the $x-y$ projection of the 3d 
particle distribution ($N=20000$) at t=25.  
In panel (a) ${\cal R}_0=0.5$, the symmetry remain unbroken, while in panel (b) ${\cal R}_0=0.01$ 
there is a spontaneous
symmetry breaking. Note
that because of the particle evaporation a 3d system does not relax to a QSS.}\label{fig3}
\end{figure}

In Fig. \ref{fig2b} we show the snapshots of two QSS to which the system relaxes after
a few oscillations.  In agreement with the theory, 
depending on the virial number, one of the QSS is spherically symmetric
while the other one is not.
For $d=3$ the integral in Eq. (\ref{gi}) cannot be performed in terms of simple analytical 
functions, and must be evaluated numerically. To locate the instability, 
we once again make use of the variables defined in Eq. (\ref{defs}) and expand 
Eq.~(\ref{enve2}) to linear order in $\Delta_i$. For $d=3$, there are two independent variables 
$\Delta_1$ and $\Delta_2$. Numerical integration of these equation shows, once again, existence of
an instability for small and large virial numbers.  To precisely locate the instability we fix 
$\epsilon_1^2=\epsilon_2^2=\epsilon_3^2={\cal R}_0$.
To linear order the dynamics of equations for $\Delta_1$ and $\Delta_2$ then decouples and becomes 
identical. This means that 
we can study the stability using a single
$\Delta(t)$ variable. The matrix that determines the evolution of the displacement 
vector from $(\Delta=0,\dot\Delta=0)$ fixed point now takes the form
\be
{\bf M}=\left( \begin{array}{cc}
0 & 1 \\
{R-15{\cal R}_0\over 5R^4} & 0\end{array} \right),
\ee
where $R(t)$ is given by Eq.~(\ref{R}) with $d=3$. 
Substituting this matrix in Eq. (\ref{m}) and adopting the procedure analogous to the one
used before, 
we find that the fixed point  $(\Delta=0,\dot\Delta=0)$, 
becomes unstable for ${\cal R}_0<0.388666...$ and ${\cal R}_0>1.61133...$.  Fig. \ref{fig3}  shows two 
snapshots of the evolution of a 3D gravitational systems.  As predicted by the theory, both for small and
large virial numbers the spherical symmetry of the initial distribution is broken by the parametric 
resonances.  
 
For 3d systems finite angular momentum can also lead to 
breaking of the spherical symmetry.  This, however, is not the case in 2d.  
Furthermore, in our simulations the initial particle distribution has very small angular momentum  --- 
in the thermodynamic limit it will be exactly zero.  
The rotation of the system is, therefore, very slow, 
while the instability happens very quickly, showing that the residual angular momentum does not play 
any role for the symmetry breaking studied in this paper.

It is interesting to compare and contrast the Rayleigh instability of charged conducting droplets and
the instability of 
self-gravitating systems.  While the Rayleigh instability is a true thermodynamic
transition, the gravitational symmetry breaking is not.  
When the charge on a droplet exceeds the critical value $Q_c$, it will undergo a first order 
transition to a prolate ellipsoid. On the other hand, the instability of a self-gravitating system is a
purely dynamical phenomenon, arising from a parametric resonance that drives an asymmetric mode of oscillation.  
The magnitude 
of the instability is saturated by 
the non-linear Landau damping~\cite{LePa08} which leads to the formation of a non-equilibrium
core-halo QSS.  If the instability occurs, the 
broken ergodicity~\cite{BeTe12} prevents the symmetry from being restored. 
In $d=2$,
a self-gravitating system with {\it finite number of particles} 
will eventually relax to thermodynamic equilibrium in which the
distribution function will have the usual Boltzmann-Gibbs form~\cite{TeLe10} and the mean-gravitational
potential will once again be spherically symmetric. The relaxation time to equilibrium, however,
diverges with $N$, so that in practice a sufficiently 
large system (such as an
elliptical galaxy)
will never evolve to equilibrium, but will stay in a non-equilibrium stationary state forever~\cite{Ma13}.  
For such systems once the instability occurs, the symmetry will remain irrevocably broken. 
This work was partially supported by the CNPq, FAPERGS, INCT-FCx, and by the US-AFOSR under the grant 
FA9550-12-1-0438. Numerical simulations have been performed at the cluster of the SIGAMM
hosted at ``Observatoire de C\^ote d'Azur'', Universit\'e de Nice --
Sophia Antipolis.

\end{document}